\documentclass[vecphys]{svmult}

\usepackage{makeidx}         
\usepackage{graphicx}        
\usepackage[bottom]{footmisc}

\begin{document}

\title*{Radio Astronomy Data Model for Single-Dish Multiple-Feed Telescopes,
and Robledo Archive Architecture}
\titlerunning{RADAMS Data Model and Robledo Archive Architecture} 
\author{J.D.~Santander-Vela\inst{1}\and
E.~Garc\'ia\inst{1}\and
J.F.~G\'omez\inst{1}\and
L.~Verdes-Montenegro\inst{1}\and
S.~Leon\inst{2}\and
R.~Guti\'errez\inst{3}\and
C.~Rodrigo\inst{3}\and
O.~Morata\inst{3}\and
E.~Solano\inst{3}\and
O.~Su\'arez\inst{3}}
\authorrunning{J.D.~Santander-Vela, E. Garc\'ia et al.} 
\institute{Instituto de Astrof\'is{\i}ca de Andaluc\'{\i}a (IAA-CSIC), Apdo.
3004, 18080 Granada, Spain \and Instituto de Radioastronom\'ia Milim\'etrica
(IRAM), Avda. Divina Pastora 7, local 20, 18012 Granada, Spain \and
Laboratorio de Astrof\'isica Espacial y F\'isica Fundamental (LAEFF-INTA),
Apdo. 50727, 28080 Madrid, Spain}
%
%
\maketitle

\vspace*{-9cm}
\noindent
{\small To appear in the accompanying CD-ROM to {\em Highlights of Spanish
Astrophysics IV}, 2007, eds.\ F.\ Figueras, J.M.\ Girart, M.\ Hernanz, C.\
Jordi (Springer)}

\vspace*{8.2cm}

\begin{abstract}
All the effort that the astrophysical community has put into the
development of the Virtual Observatory (VO) has surpassed the non-return
point: the VO is a reality today, and an initiative that will self-sustain,
and to which all archival projects must adhere. We have started the design
of the scientific archive for the DSS-63 70-m antenna at NASA's DSN station
in Robledo de Chavela (Madrid). Here we show how we can use all VO proposed
data models to build a VO-compliant single-dish, multiple-feed, radio
astronomical archive data model (RADAMS) suitable for the archival needs of
the antenna. We also propose an exhaustive list of Universal Content
Descriptors (UCDs) and FITS keywords for all relevant metadata. We will
further refine this data model with the experience that we will gain from
that implementation.
\end{abstract}

\section{Introduction}\label{secIntroduction}

The AMIGA project (Analysing the interstellar Medium of Isolated GAlaxies)
was born in 2003, and intends to provide a statistical characterisation of a
strictly selected sample of isolated galaxies composed by more than 1000
objects, by means of multi-wavelength data, and with a particular emphasis
on radio data at cm, mm, and sub-mm wavelengths. All these data are being
periodically released via the web page of the
project\footnote{\texttt{<http://www.iaa.csic.es/AMIGA.html>}}, which will
soon provide a Virtual Observatory-compliant interface.

AMIGA+ is the natural extension to AMIGA, with three different goals:
exploitation of the AMIGA catalog selecting the best candidates for a
detailed study of isolated galaxies, scientific extension to the millimeter
and submillimeter range, and participation in the development of systems
allowing the access and display of large radio astronomical databases, both
single-dish and interferometric.

\subsection{Radio Astronomy in the VO}\label{ssecVO}

The VO can be defined as a set of protocols and data models that allow for
easy discoverability of interoperable data-sets, which share an unified
description by means of a common data model.

The protocols define how data are archived, searched for, and accessed,
while data models describe the set of entities needed for information
storage in a particular field, and specify both the data being stored, and
the relationships between them. By establishing generic relationships, a
data model able to store information for many different instruments can be
defined, but in order to set the parameters for each antenna each telescope
and instrument has to be separately studied.

Within the VO, the common interchange format is the VOTABLE, a
FITS~\cite{Hanisch:1999fk} replacement based upon XML \cite{Quin:2006fk},
while the communication protocols (such as the Simple Spectral Access
Protocol, SIAP \cite{Dolensky:2006fk}) are based upon common web-services
technologies (SOAP \cite{Mitra:2003uq}, XML-RPC \cite{Winer:1999kx}).

These common standards have been possible by the joint work of an
international standardisation body, the International Virtual Observatory
Alliance (IVOA) \cite{Hanisch:2003vn}. IVOA is a federation of national and
supranational VO groups, and steers and sanctions the development of the
different parts of the VO infrastructure, thanks to its different Working
Groups. In particular, the Data Access Layer and Data Modelling groups are
the ones trying to standardise access protocols and data models within the
VO, such as the already mentioned SSAP, or the Space and Time Coordinates
(STC) data models, and many others.

However, there is only one radio-related data model proposal within the
IVOA, Lamb and Power's \emph{IVOA Data model for raw radio telescope data}
\cite{LamPow0310IVOA}, but it is prior to many IVOA data modelling
developments, and it constitutes a proposal for radio interferometry. Some
of our efforts have been inspired by this proposal.

In a search for VO-compliant radio archives we only found the \emph{ATCA
(Australian Telescope Online Archive) Data Model} \cite{2006astro.ph..1354}.
Again, even when it provides VO interfaces, it makes use of very few
standard data models, and relies in custom implementations instead.

As no suitable data model existed for single-dish radio astronomy, and in
order to develop a VO-compliant archive for single-dish antennas, while
contributing to the development of the radio VO, a complete radio data model
had to be defined.

\section{RADAMS: Radio Astronomy DAta Model for Single-dish radio
telescopes} \label{secRADAMS}
Our goal is to develop a data model suitable
for storage of single-dish spectroscopic data, in order to develop the
scientific archive for the DSS-63 antenna, as a result of our collaboration
with the LAEFF-INTA .

The DSS-63 is a very sensitive 70-m antenna, part of NASA's Deep Space
Network (DSN), located in the Madrid Deep Space Communications Centre
(MDSCC) at Robledo de Chavela, and its main use is the monitoring and remote
command of NASA's missions in the Solar System. However, when not performing
DSN-related tasks, the DSS-63 can make use of its 22MHz K-band receiver to
perform spectroscopic observations with a 2 to 16 MHz bandwidth digital
(384-samples) spectrometer. Observing time is allocated by LAEFF-INTA by
agreement with NASA, with up to 260h reserved for host-country astronomers.

\subsection{Data Model Overview}\label{ssecDataModelOverview}

The data model that we have developed is specific to single-dish radio
telescopes, and can hold data from multiple instruments for a single
antenna, as long as they are spectrometers.

In order to build the RADAMS, we have reviewed the different data models
offered by the IVOA, selected the most suitable for our needs, and then
decided how to better group those data models into a single one.

Three are the main references for our data model:

\begin{itemize}
	\item Data Model for Observation (DMO) \cite{McDBonGia0405Data}
	
	\item Data Model for Astronomical Dataset Characterisation (DMAC)
	      \cite{McDBonChi0605Data}
	
	\item Spectral Data Model (SDM) \cite{McDowell:2006fk}
\end{itemize}

\begin{figure}
\centering
\includegraphics[width=\columnwidth]{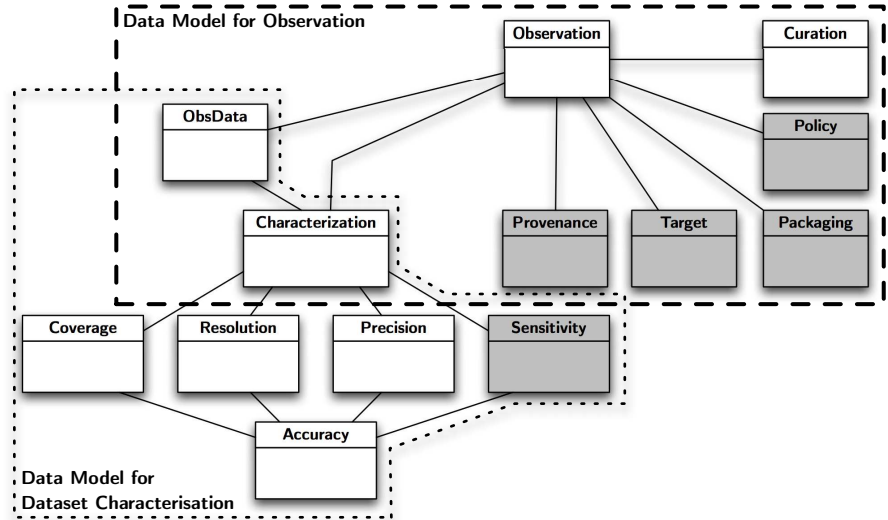}
\caption[RADAMS' components]{
	RADAMS' component classes and their origin. In white, classes used as
specified by IVOA; in gray, classes specified and/or modified by the RADAMS.
}
\label{figRADAMS}       
\end{figure}

The main components of the RADAMS, together with the documents inspiring
each section, are shown in Fig. \ref{figRADAMS}. In gray we have marked the
components that needed further specification, because they have not been
developed by current IVOA standards, or whose definition had to be changed
in order to accommodate RADAMS' specifications. The SDM is only used in the
data import/export sequence.

In particular, Packaging and Policy had to be fully specified, and for our
packaging and data delivery needs we have developed the VOPack, a system for
fully characterising and delivering VO-compatible datasets. As for the
Policy class, a role based mechanism will be provided, that allows for
different data access levels for different \emph{roles}, instead of basing
permissions upon users.

More than 40 classes and subclasses are specified by the RADAMS. For each
class, the RADAMS specifies an \textbf{Attribute name}, the corresponding
{FITS Keywork} (both for importing and exporting data), the associated
\textbf{UCD} (Universal Content Descriptor: gives semantic information about
the attribute, and allows for attribute matching between databases with
different attribute names), and finally a full \textbf{Description} of the
attribute, explaining the meaning and possible values of each attribute.

\subsection{Archive Infrastructure}\label{ssecArchiveInfrastructure}

\begin{figure}
\centering
\includegraphics[width=\columnwidth]{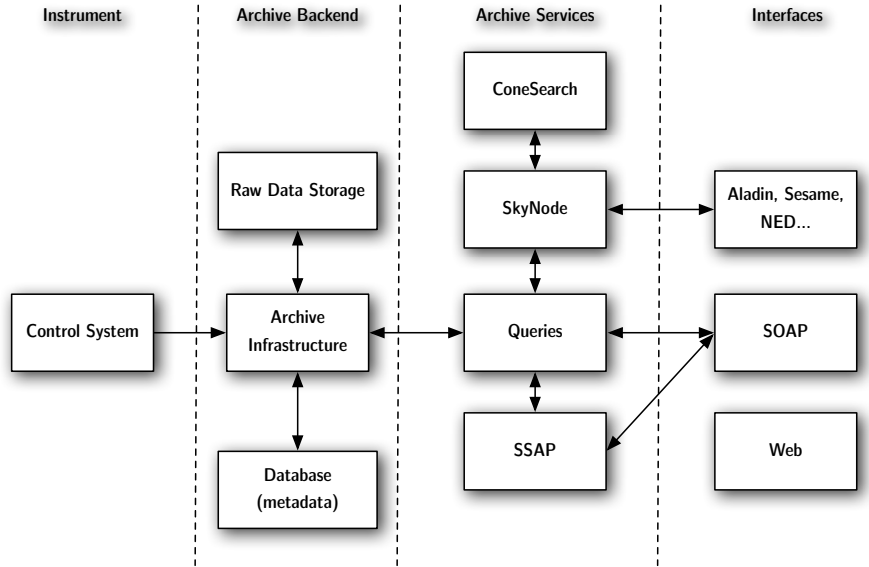}
\caption[Robledo Archive Infrastructure]
	{Layered Robledo Archive infrastructure.}
\label{figArchiveInfrastrure}       
\end{figure}

The different archive components have been distributed in different layers
in order to minimize layer and component communication, which allows for
easier development of the archive infrastructure. The different layers and
components are shown in Fig. \ref{figArchiveInfrastrure}.

From left to right, the different layers contain software packages, running
from the Instrument (the Control System) to the Interfaces for users or
software packages performing queries to the archive.

The telescope's Control System writes data, and after being processed by a
semi-automated workflow, the Archive Backend will store both the raw data
and the workflow-generated metadata. The Archive Services layer will allow
several components to perform queries on the stored data and metadata, and
provides the infrastructure for the Interfaces layer, where user- or
machine-operated clients will be able to use standard VO protocols to query
the archive.

\section{Conclusions and Future Work}\label{secConclusions}

The IVOA modelling efforts have built a solid foundation upon which we have
built the RADAMS. However, we needed to perform a careful selection and
synthesis of the provided data models in order to provide a unified data
model for the DSS-63.

In particular, the RADAMS provides definitions for the Provenance, Policy
and Packaging classes, with the addition of the VOPack, a VO-specific
XML-based packaging and delivery format.

We plan to release the RADAMS as an IVOA Note for the IVOA Radio interest
group, and present the VOPack to the IVOA Data Access Layer Working Group
for evaluation.

At the same time, we will implement the RADAMS as the foundation for the
scientific archive of the DSS-63 antenna, and we will extend its
capabilities in order to be able to use the RADAMS with additional antennas.

\section*{Acknowledgements}

This work has been partially supported by DGI Grant AYA 2005-07516-C02-01
and the Junta de Andaluc\'ia (Spain). JDSV acknowledges support at the
IAA/CSIC by an I3P Postgrade Grant, funded by the European Social Fund. The
IAA-LAEFF collaboration is partially funded by the Spanish Virtual
Observatory Network, funded by the PNAYA, AYA2005-24102-E.

%
%
\bibliographystyle{ieeetr}
\bibliography{obs_talk_santander-vela_1.bib}
%


\end{document}